# Efficient broadband sound absorption exploiting rainbow labyrinthine metamaterials

F. Nistri[1, 2], V. H. Kamrul[2], L. Bettini[2], E. Musso[2], D. Piciucco[2], M. Zemello[2], A.S. Gliozzi[1], A.O. Krushynska[3], N. M. Pugno[4,5], L. Sangiuliano[6], L. Shtrepi[7], and F. Bosia[1, †]

[1] Department of Applied Science and Technology, Politecnico di Torino, Torino, Italy

[2] Politecnico di Milano, Milano, Italy

[3] Faculty of Science and Engineering, University of Groningen, Groningen, The Netherlands

[4] Laboratory for Bioinspired, Bionic, Nano, Meta Materials & Mechanic, University of Trento, Trento, Italy

[5] School of Engineering and Materials Science, Queen Mary University of London, United Kingdom

[6] Phononic Vibes s.r.l., Milano, Italy

[7] Department of Energy "Galileo Ferraris", Politecnico di Torino, Torino, Italy

†Corresponding author: federico.bosia@polito.it

## Abstract

In this work, we demonstrate in a proof of concept experiment the efficient noise absorption of a 3-D printed panel designed with appropriately arranged space-coiling labyrinthine acoustic elementary cells of various sizes. The labyrinthine unit cells are analytically and numerically analysed to determine their absorption characteristics and then fabricated and experimentally tested in an impedance tube to verify the dependence of absorption characteristics on cell thickness and lateral size. The resonance frequency of the unit cell is seen to scale approximately linearly with respect to both thickness and lateral size in the considered range, enabling easy tunability of the working frequency. Using these data, a flat panel is designed and fabricated by arranging cells of different dimensions in a quasi-periodic lattice, exploiting the acoustic "rainbow" effect, i.e. superimposing the frequency response of the different cells to generate a wider absorption spectrum, covering the target frequency range, chosen between 800 and 1400 Hz. The panel is thinner and more lightweight compared to traditional sound absorbing solutions and designed in modular form, so as to be applicable to different geometries. The performance of the panel is experimentally validated in a small-scale reverberation room, and an absorption close to ideal values is demonstrated at the desired frequencies of operation. Thus, this work suggests a design procedure for noise-mitigation panel solutions and provides experimental proof of the versatility and effectiveness of labyrinthine metamaterials for tunable mid- to low-frequency sound attenuation.

## 1- Introduction

In recent years, Acoustic Metamaterials (AMs) have gained widespread attention because of their exceptional properties, which are not commonly found in naturally occurring materials [1–3]. AMs could potentially pave the way to the development of a new generation of acoustic absorbers and diffusers with deep-subwavelength thickness, that can be tailored for a desired frequency spectrum [4]. Their use brings new possibilities to the traditional problem of achieving low-frequency absorption [5]. In addition, AMs offer the possibility to achieve high performance in terms of noise reduction, and simultaneously reduced size and weight of structures [6], going beyond the limitations of conventional technologies based on single layer mass law, double layer resonance frequency tuning and porous absorber thickness optimization [7]. In particular, these new materials seem to be promising and to respond to the thickness and weight constraints imposed by the design/technological requirements of the market, e.g. in aircraft cabin design in aeronautics [8]. AMs can be combined with conventional solutions like porous materials [9], Helmholtz resonators [10] or tensioned membranes [11,12] for tuned or optimized performance. It is well known that perfect absorption can be obtained when a critical coupling condition occurs, whereby thermoviscous losses are exactly balanced by the energy leakage [13]. Such perfect absorption in a subwavelength regime has for example been shown to be achievable with periodic arrays of vertical Helmholtz resonators [14], as well as with plate-resonator/closed waveguide structures [15]. However, the working frequencies of these AMs are often quite narrow, or the structures need to be bulky to enable broadband operation. To address this problem, the concept of "rainbow trapping" in acoustic resonators with variable parameters, and hence working frequencies, [16,17]or systems with asymmetric porous absorbers [18] has been adopted so far.

A particularly interesting type of AM that has emerged in recent years are "labyrinthine" or "coiled" structures [19]. These are based on exploiting acoustic wave propagation in curved channels of subwavelength cross-section, giving rise to an extremely high effective refractive index (and thus to a decrease of the effective wave speed) and the possibility of achieving "double negativity", i.e.

simultaneously negative effective density and bulk modulus, or conical dispersion [20]. Tapered 2-D labyrinthine designs have also been shown to achieve optimal broadband impedance matching, which is fundamental for efficient absorption [21]. The concept has also been extended from 2-D to 3-D space coiling labyrinthine structures [22]. Experimental demonstrations of the theoretically predicted broadband negative refractive index have been achieved through reflection or transmission measurements and two-dimensional prism-based measurements on 3-D printed thermoplastic labyrinthine samples [23]. Hilbert-like fractal acoustic metamaterials have also been designed, fabricated via 3-D printing, and experimentally characterized, to achieve efficient low-frequency acoustic wave attenuation [24,25] . 3-D "one-port" labyrinthine structures have also been proposed to achieve high-levels of sound absorption over large frequency ranges (and for various incidence angles), exploiting different channel lengths to tune the operating bands [26]. Another example of labyrinthine AM providing large tunability are spider-web inspired structures, in which the addition of edge cavities can further enhance the possibilities to manipulate dispersion properties, control the appearance of band gaps or negative group velocity, and tailor transmission/reflection characteristics [27]. Several studies have shown that the use of space filling structures, such as Wunderlich curves, can efficiently control transmission, reflection and absorption by varying channel tortuosity, so that total broadband reflection/absorption can be achieved e.g. by tuning the channel length [28,29].

Thus, labyrinthine and space-filling AMs have provided a very convenient and efficient way to achieve sound control in large frequency ranges, especially in the subwavelength regime, by tuning geometrical design parameters (e.g., channel tortuosity or elongation and cavity size). This kind of adaptability could largely benefit noise absorption applications at small to medium scale, where restrictions on structural size of the absorbers impose trade-offs between efficiency and encumbrance. Sufficiently thick conventional acoustic absorbing materials, such as glass wool or foams within sandwich panels with an average density of 75 kg/m3 [30] can absorb acoustic wave energy in wide frequency ranges, but their bulky characteristics limit their broad application for low frequency absorption. Moreover, lightweight characteristics become crucial when dealing with devices in the aerospace and automotive industry or other technological domains [8]. At present, few studies [31] in the literature have presented detailed acoustic characterization studies on large structures such as metamaterial-based panels, and none on labyrinthine ones, to the best of our knowledge. Further, the need emerges to investigate labyrinthine AM performance on structures that are closer to potential operating conditions, i.e. in diffuse field conditions.

With this in mind, the present study investigates the design of coiled labyrinthine acoustic resonators, their appropriate combination in a 3-D printed “rainbow” panel, and presents a proof-of-concept experiment to demonstrate efficient broadband noise absorption. The paper describes the following workflow.Design of labyrinthine Unit Cells (UCs) and analytical model of their absorption spectra is described in Section 2.Experimental and numerical characterization and comparison with the analytical model for the designed UC is provided in Section 3.Design and 3-D printing of a full-scale panel with variable, appropriately chosen constituent UC dimensions, its numerical model and experimental test in a reverberation room is described in Section 4. Finally, a numerical investigation of a coupling of the panel with conventional absorbing material is described in Section 5.

## 2 Unit cell design and analysis

In conventional ideal fluid environments, sound propagation is typically described by a lossless wave equation [32–34]. However, when fluids are confined in small regions, like narrow tubes or slits, it becomes necessary to account for structural losses [30,33,34]. There are two primary loss mechanisms: thermal losses, primarily arising from heat diffusion at the boundaries within the sound field, and viscous losses, due to friction with the UC walls. These physical dissipation mechanisms effectively absorb sound energy, leading to perfect absorption when they are balanced by the acoustic energy leakage by radiation from the UC: the matching of the two corresponds to the critical coupling condition [33,35].

The type of labyrinthine geometry considered in this work (Fig. 1a) is a first-iteration Wunderlich curve [36], which corresponds to a quarter-wave resonator. This design can be studied using narrow tube thermo-viscous loss models to derive its absorption properties. In particular, the designed aperture (Fig. 1a) is responsible both for the losses and energy leakage of the UC, while the cavity (Fig. 1b) is only responsible for the losses. The considered UC can be modelled as a tube with constant cross section $w \times t$, and a rigid backing (Fig. 1c) as pointed out by several works [37–39]. In particular, given the subwavelength size of the system, the equivalent tube can be described as a narrow duct, following the Stinson's approximation for thermo-viscous losses [40], while the inlet can be studied as a perforated plate using the Johnson-Champoux-Allard (JCA) model, as described in [37].

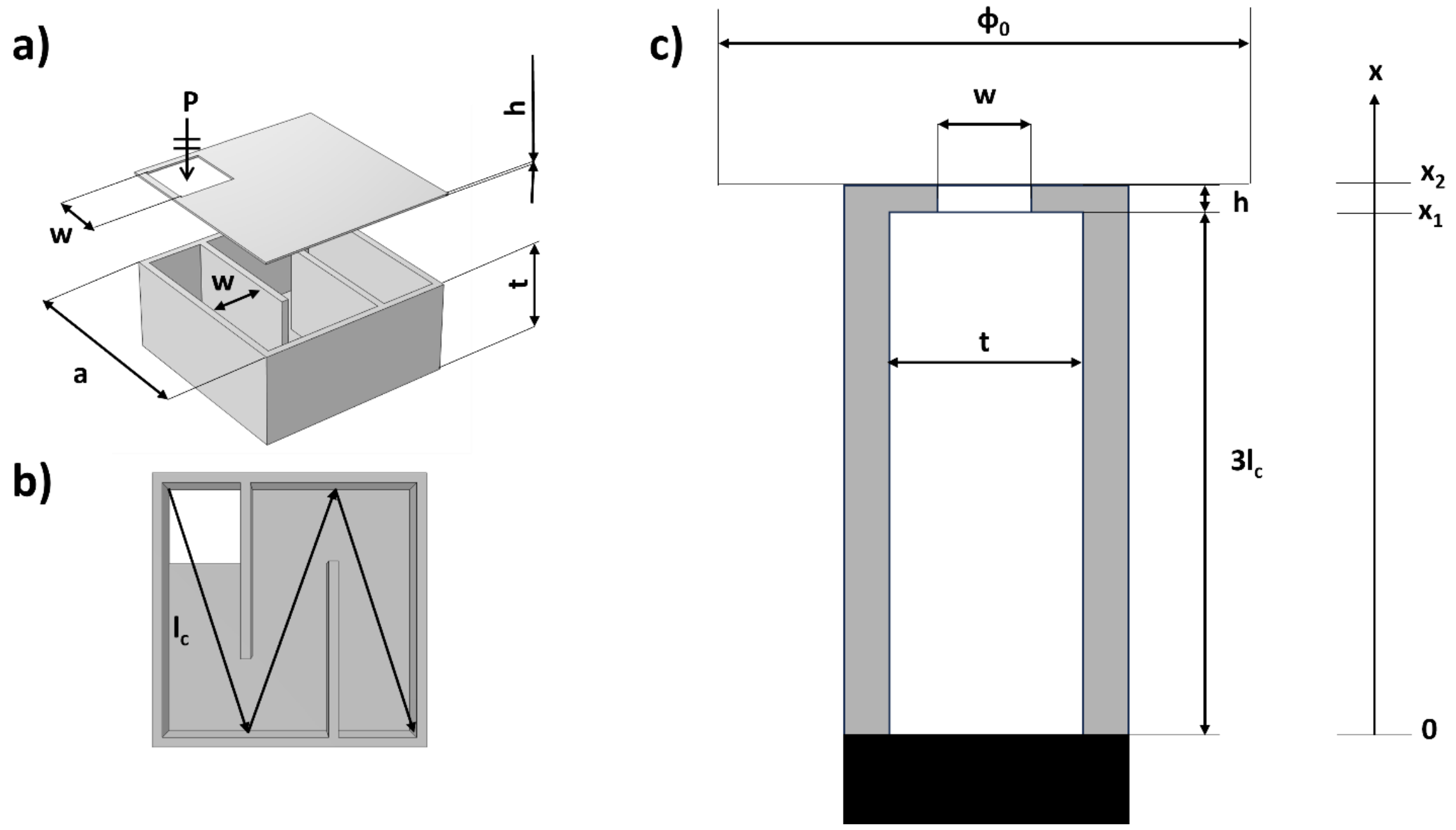

*Figure 1: (a) 1st iteration Wunderlich curve UC with w=6 mm, t=9.6 mm, a=19.6 mm and h=0.5 mm. (b) Schematic of the effective distance travelled by the pressure wave inside the cavity. (c)Schematic of the equivalent model with the tube of variable cross section closed by a reflecting backing and open on one side.*

Thus, calculating the impedance of the equivalent model at the surface ($Z_s$), the acoustic absorption ($\alpha$) can be computed as:

$$\alpha = 1 - \left|\frac{Z_s - Z_0}{Z_s + Z_0}\right|^2 , \tag{1}$$

where $Z_0 = \rho_0 c_0$, $\rho_0$ =1.225 kg/m$^3$ are the impedance and air density respectively and $c$ =343 m/s is the speed of sound in air [29]. The value $Z_s$ and its derivation are provided in Appendix I.

A parametric study of the analytical model is carried out (Fig.2) to study the dependence of maximum sound absorption $\alpha_{max}$ on the edge size $a$ (scaling the UC isotropically in the plane), ranging from 10 to 29 mm, and the aperture cross section $S_a$ from approximately 22 to 40 mm$^2$. The thickness $t$ = 9.6 mm is fixed to obtain a subwavelength thickness in the investigated frequency ranges. From the four-parameter map obtained, a UC featuring perfect absorption in the targeted frequency range ($a$ = 19.7 mm and $S_a$ = 36 mm$^2$) is selected. The cross section $S_a$ is chosen so as to allow measurements in an impedance tube (see Section 3).

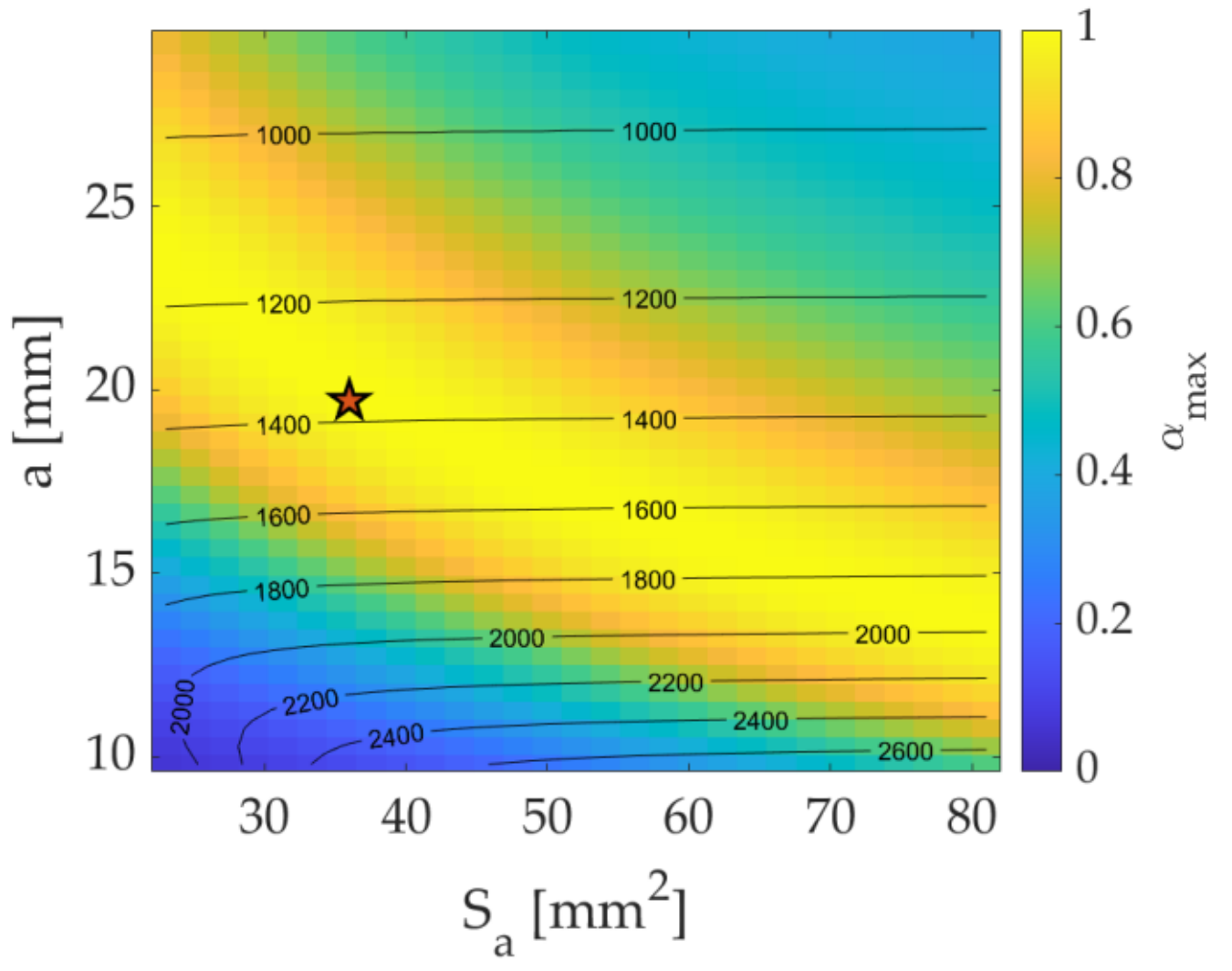

*Figure 2: Maximum absorption map vs the edge size a vs the aperture cross section $S_a$ of the UC. The contour lines represent the frequencies of maximum absorption, and the star is the selected UC size.*

The analytically calculated absorption spectra of the UC (Fig.3a) confirm that perfect absorption is attained at 1370 Hz with a peak width of $\Delta f = 110$ Hz. In addition, to gain further insight on the absorption phenomenon, the reflectance $|r|^2 = 1 - \alpha$ of the UC is represented in the complex frequency plane using the method introduced by Romero Garcia et al. [35]. As shown in Fig. 3b in the lossless case, the reflectance shows a zero/pole complex conjugate pair attained at the resonance frequency of the UC. When losses are introduced in the model, the pair is shifted towards smaller imaginary frequencies and the critical coupling condition [35] is obtained when the losses balance the energy leakage. In this case, the zero falls exactly on the real frequency axis, resulting in perfect absorption (Fig. 3c).

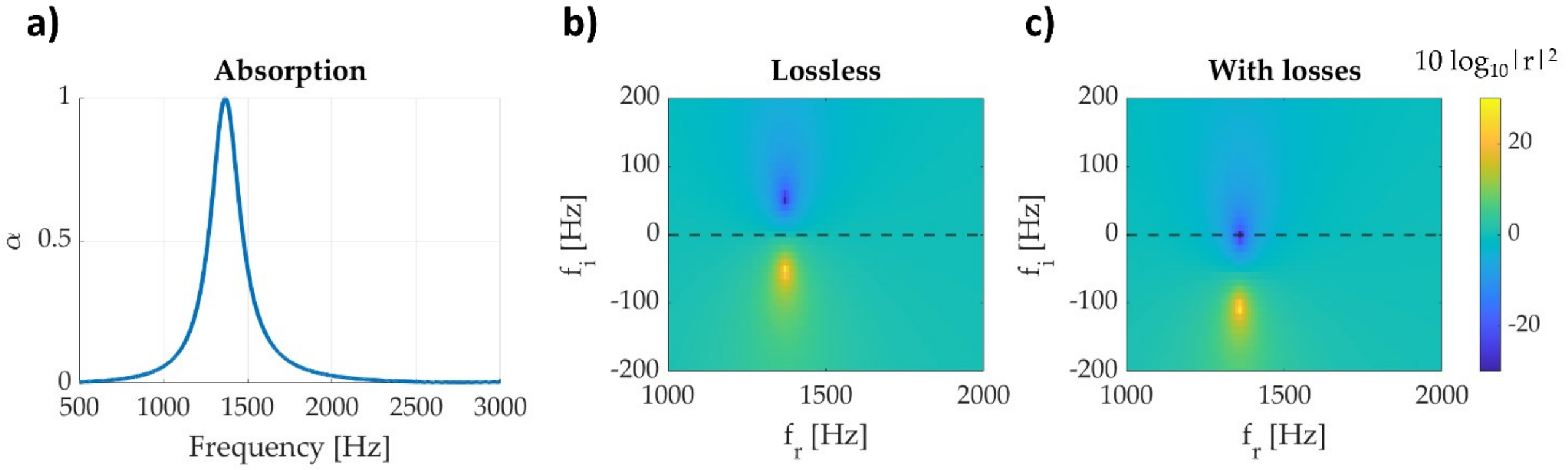

*Figure 3: (a) absorption spectrum for the 1st iteration Wunderlich curve UC. (b), (c) Reflectance of the UC represented on the complex frequency plane for the lossless and with losses case, respectively.*

## 3 Unit cell experimental and numerical characterization

In order to validate the design predictions and experimentally assess the efficiency of the UC subwavelength sound wave absorption, we first perform normal incidence measurements on various types of UCs in an impedance tube. The UCs consist of a square cavity inside a solid cylinder, in which internal solid walls form a labyrinthine pattern. The UC samples for impedance tube measurements are manufactured in PLA using a Dremel Digilab filament 3D printer and in VeroYellow using a Stratasys J750 FDM 3D printer. The two polymers have similar mechanical characteristics, which are compatible with the rigid wall hypothesis used in simulations.

The impedance tube setup is depicted in Fig. 4. The impedance measurements are performed in accordance with ISO 10534-2 [41] and ASTM E1050-19 [42] (two-microphone technique) regulations, in order to measure the normal incidence absorption coefficient ($\alpha_0$), calculated for a normally impinging pressure wave [43]. Measurements are performed using a HW-ACT-TUBE and -STL (Siemens, Munich, Germany), which has a diameter of 35 mm and is equipped with two 1/4 " flush mounted GRAS 46BD microphones (GRAS, Holte, Denmark) and a white noise source, i.e., a 2" aluminium driver. The method allows to obtain accurate sound pressure amplitude and phase measurements in the whole frequency range of interest, i.e., 100-5000 Hz [43], with a spectral resolution of 2 Hz. The sample holder consists of a tube with a sliding rigid piston. A specific plastic sample holder is used as shown in Fig. 4b. The method allows a better control on the positioning of the different samples into the flanged sample holder of the tube. The 3D printed samples have a reflective overall surface and a diameter allowing for tolerances such that the circumferential effect discussed in [44] can be considered negligible. The effect of any possible irregularity in the samples, and in particular at the edges, is taken into consideration by repeating the tests for each structure with three nominally identical samples, to evaluate reproducibility. Temperature and atmospheric pressure are monitored by calibrated transducers.

a) b)

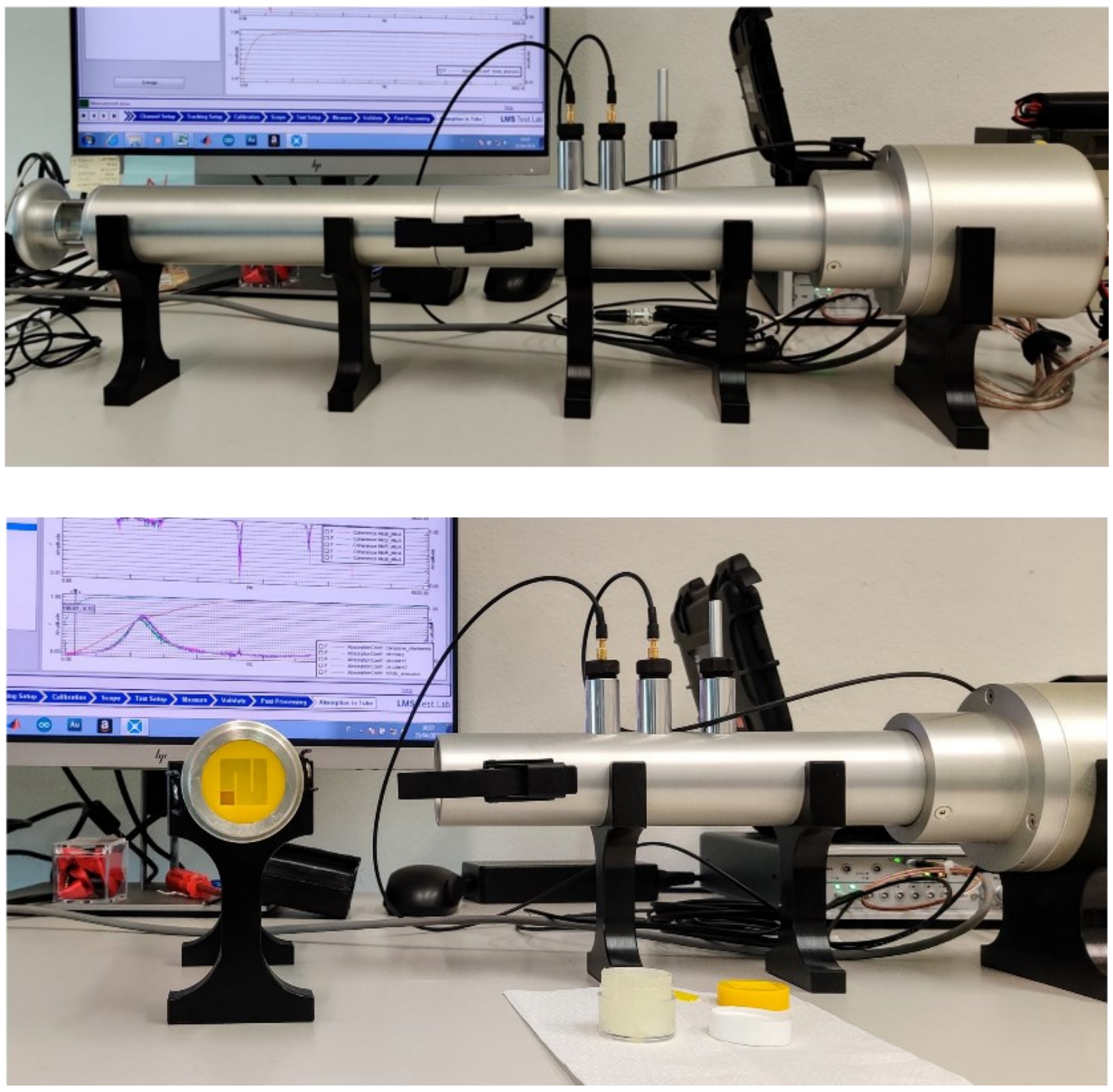

*Figure 4: Experimental setup for impedance tube measurements. (a) The complete impedance tube setup. b) Opened tube with sample mounted in its holder.*

Repeated measurements are performed on each UC, to assess reproducibility. The acoustic pressure is measured by two microphones set at a distance of 29 mm from each other along the length of the tube. The absorption coefficient spectrum $\alpha(f)$ of the UC is obtained using the transfer function method [32]. This procedure is repeated several times for each experimental configuration to reduce dispersion in the measurements.

For further comparison, a numerical Finite Element (FE) model reproducing the experimental setup is implemented, as shown in Fig. 5a, comprising the air domains of the tube, the aperture, and the UC. The model is implemented in COMSOL Multiphysics, and is performed using the Acoustics module and “narrow region” condition for the tube and the UC, and “poro-acoustic” one for the aperture. At the boundary located at the position $x_0$ a plane wave radiation of the form $P = Ae^{ikx}$, $A = 1$ Pa, $k =$ $2\pi f/c$ , where $f$ is the frequency, while the average pressure at the cross section of the tube is sampled at the surfaces located at $x_1$ and $x_2$ to compute the transfer function, and the absorption.

A comparison of the analytical, numerical, and experimental absorption spectra of one of the labyrinthine UC is shown in Fig. 5b, as an illustrative example. The tested sample has a UC lateral size of $a$ = 19.7 mm and a thickness of $t$ = 9.6 mm. The experimental absorption spectrum features an absorption coefficient of $\alpha$ = 0.95, at a frequency $f_{exp}$~1480 Hz with a peak width of $\Delta f_{exp}$ ~390 Hz. This result is used to calibrate the analytical model's correction factors $\psi = 8.4 \times 10^{-2} L$ and $\alpha_s = 1.9$ (see Appendix). Although excellent matching is achieved in the neighbourhood of the absorption peak frequency, the analytical curve underestimates the absorption for the other frequency ranges and the width of the peak by ca. 70 Hz. This underestimation is probably due to the missing thermo-viscous losses taking place on the walls of the tube itself, which are instead present in the FE simulation. In the numerical model, although the peak frequency is underestimated by 40 Hz, the spectrum presents a larger absorption for lower and higher frequencies with respect to the peak and accurately predicts the peak bandwidth $\Delta f_{num}$~360 Hz obtained in experimental results, together with the correct behaviour in the higher frequency range.

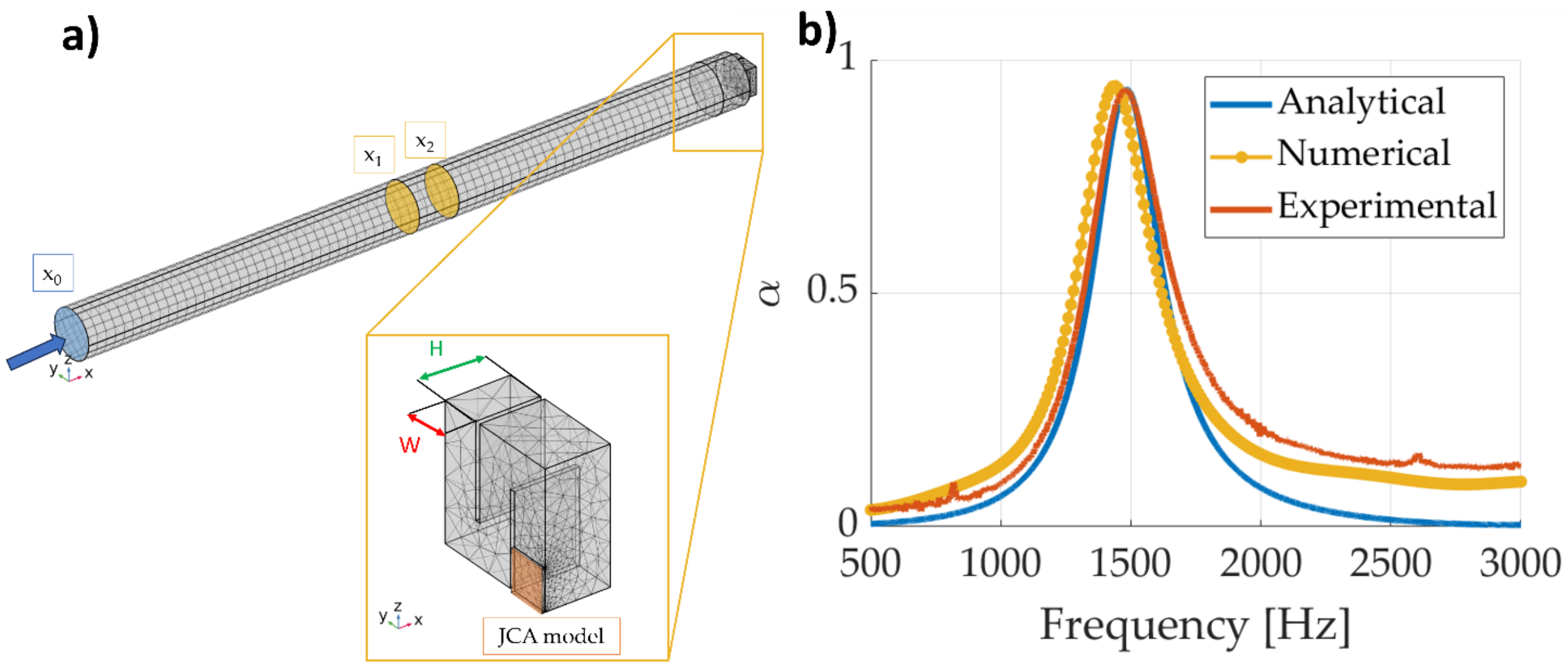

*Figure 5: a) FE mesh in the model of the impedance tube containing the labyrinthine UC. Blue transparency region corresponds to the area of application of the boundary pressure, while the yellow ones to the areas where the average pressure is sampled. In the inset, the orange transparency region*

*highlights the domain where the poroacustic conditions (JCA in analytical model) are applied and the cross-section sizes of the UC used for the narrow region module. b) Comparison of the analytical, numerical, and experimental absorption spectra.*

With the ultimate purpose of the design of the rainbow panel, the resonance frequency dependence on the dimensions of the UC cavity is studied [21]. In order to quantify the resonance frequency dependence on the geometrical size, several UCs similar to the ones in Fig. 1 were manufactured, varying both their thicknesses $t$ (within the 10.0-25.0 mm range, keeping their UC edge size $a$ fixed to 21.2 mm) and their lateral edge size $a$ (within the 13.0-20.0 mm range, keeping their thickness $t$ fixed to 12.6 mm). Numerical and experimental results in Fig. 6a and 6b show the frequency peak value vs $t$ and $a$, respectively, obtained from the impedance tube measurements and an eigenfrequency analysis of the numerical lossless model of the studied UCs using the FE model shown in Fig. 5a. The resonance frequency scales approximately linearly by a factor $\frac{df_r}{dt} = -42.7$ Hz mm$^{-1}$ with the sample thickness $t$, and by a factor $\frac{df_r}{da} = -145.3$ Hz mm$^{-1}$ with the UC edge size $a$. Thus, the influence of the UC lateral edge size allows to tune the working frequency in a larger interval, while the thickness dependency can be employed to fine-tune the frequency in a smaller frequency range to the desired value.

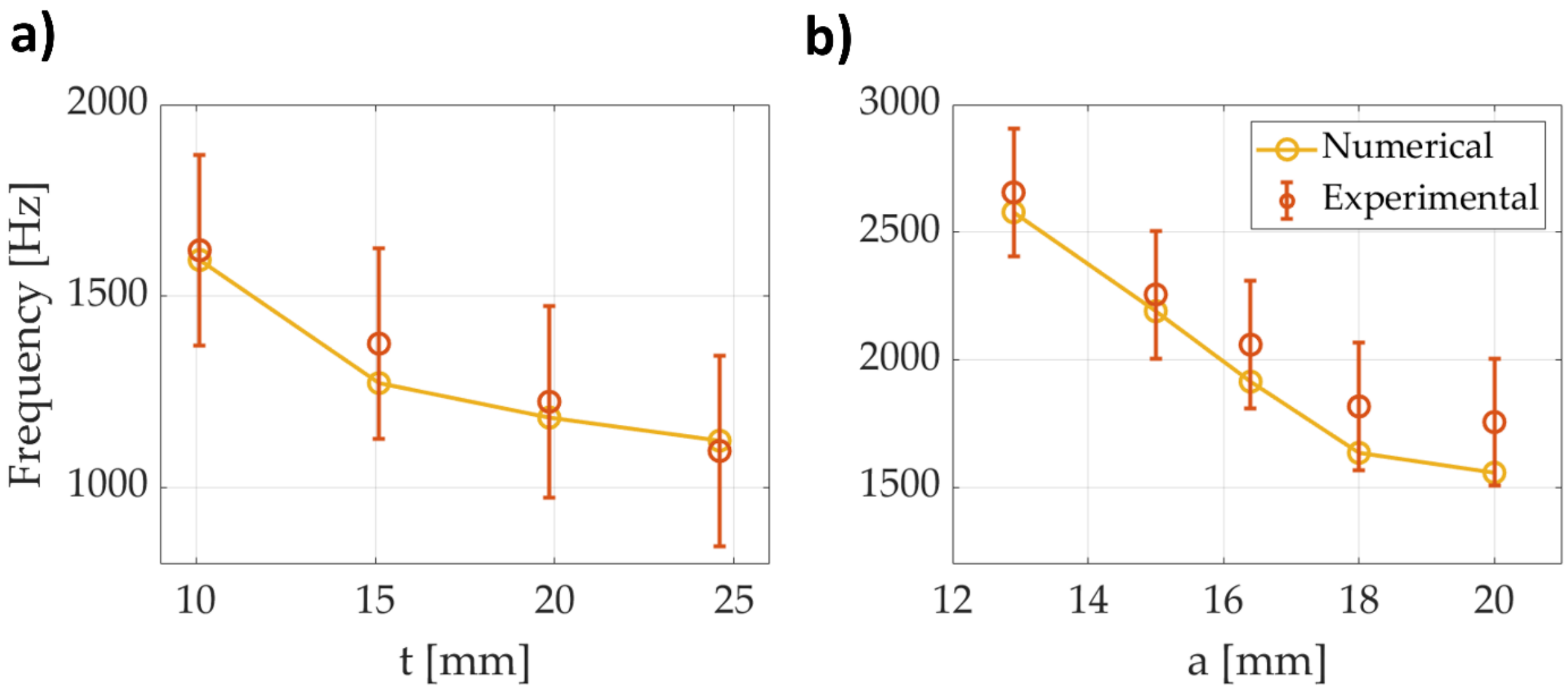

*Figure 6: Absorption peak frequency variation vs. the (a) UC thickness t and (b) UC edge size a. The error bars on the experimental results are calculated as the peak width.*

## 4 Rainbow AM labyrinthine panel

### *4.1 Panel design and fabrication*

Using impedance tube data, relative to the characterization of the chosen UC, we establish the target frequency range of interest for sound absorption (e.g., for aeronautics applications), between 800 and 1300 Hz. This can be achieved by exploiting various UCs of different lateral sizes and thicknesses, so that the resulting resonance spectrum is given by their combination, and the corresponding absorption range extends over the desired range. An AM panel is thus designed, consisting of differently sized UCs, with varying thicknesses and surface areas. The internal wall thickness of each UC is proportional to the UC edge area. Starting from two UC edge sizes ($a_1$ = 19.5 mm and $a_2$ = 26.0 mm) at different thicknesses ($t_1$ = 10÷15 mm for the $a_1$ UC, and $t_2$ = 23÷25 mm for the $a_2$ UC), a spatial pattern of repeated $a_1$ and $a_2$ UCs is conceived, covering a total module of 156 x 156 mm$^2$ area. We define this module "macro-cell" (*M* and *M'* in Fig. 7a and 7b) because of its intermediate size between an UC and the complete sound absorbing panel. The complete panel consists of the combination of 20 macro-cells arranged in the plane (Fig. 7b), to cover a total surface of 628 x 784

mm$^2$, i.e., the area of the panel. Adjacent macro-cells are rotated by 90 degrees and reflected (Fig. 7), to ensure maximum spatial homogeneity in the sound absorbing properties.

The polyamide panel is then manufactured through selective laser sintering (SLS), a technique where each level is sintered by a laser beam, directed by a scanning system. The process is repeated, and the prototype is built layer by layer. SLS is chosen as the most appropriate fabrication technique since it does not require supports to sustain oblique and horizontal protrusions and it allows to fabricate undercuts. However, to manufacture hollow structures like the UCs, it is necessary to include at least two apertures (on the top and bottom surfaces) for each UC cavity, in order to remove any residual powder from the fabrication process, so an additional opening is added for each UC. For the measurement phase, these additional openings are closed with properly sized polyamide caps that are fixed with aluminium tape to guarantee an acoustic seal. A 3D view of the final panel is shown in Fig. 7c.

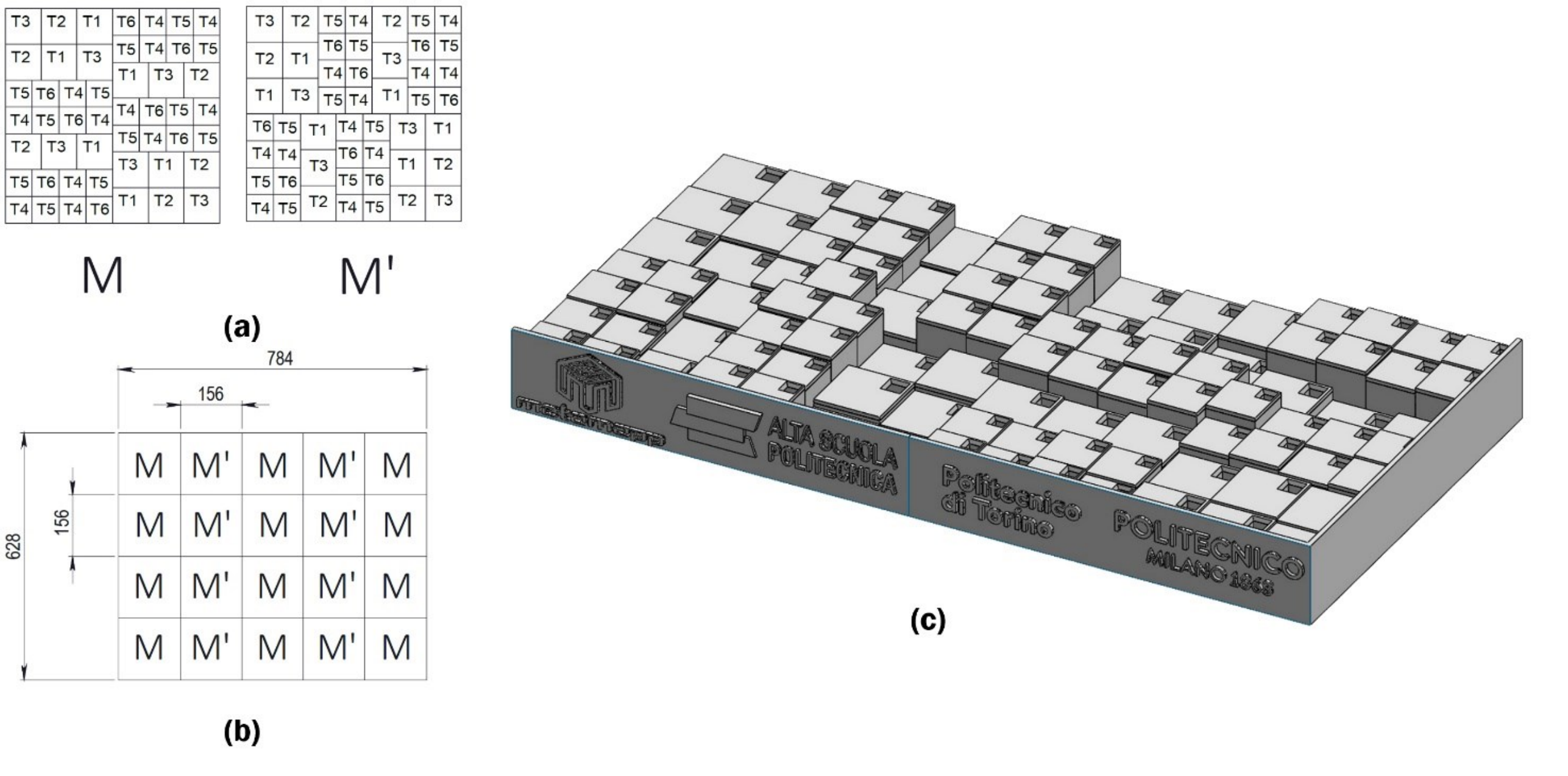

*Figure 7: Modular elements of the complete sound absorption panel. (a) Two adjacent macro-cells. Each macro-cell consists of several UCs of edge size $a_1$ and $a_2$ with varying thicknesses (T1-T6); (b) Schematic top view of the final panel, consisting of a regular arrangement of M and M' macrocells (geometrical sizes are indicated in mm); Adjacent macro-cells are reflected and rotated by 90° in*

*order to guarantee a globally homogeneous spectral response. (c) 3D view of two adjacent macrocells.*

*4.2 FE model of the AM panel*

The AM panel can be modelled in FE using six different domains, shown in Fig.8, three of which are structural domains (the AM panel side walls, the front and back surfaces) and three are acoustic domains (the UC cavity voids, and the aperture "caps" on the front and back surfaces).

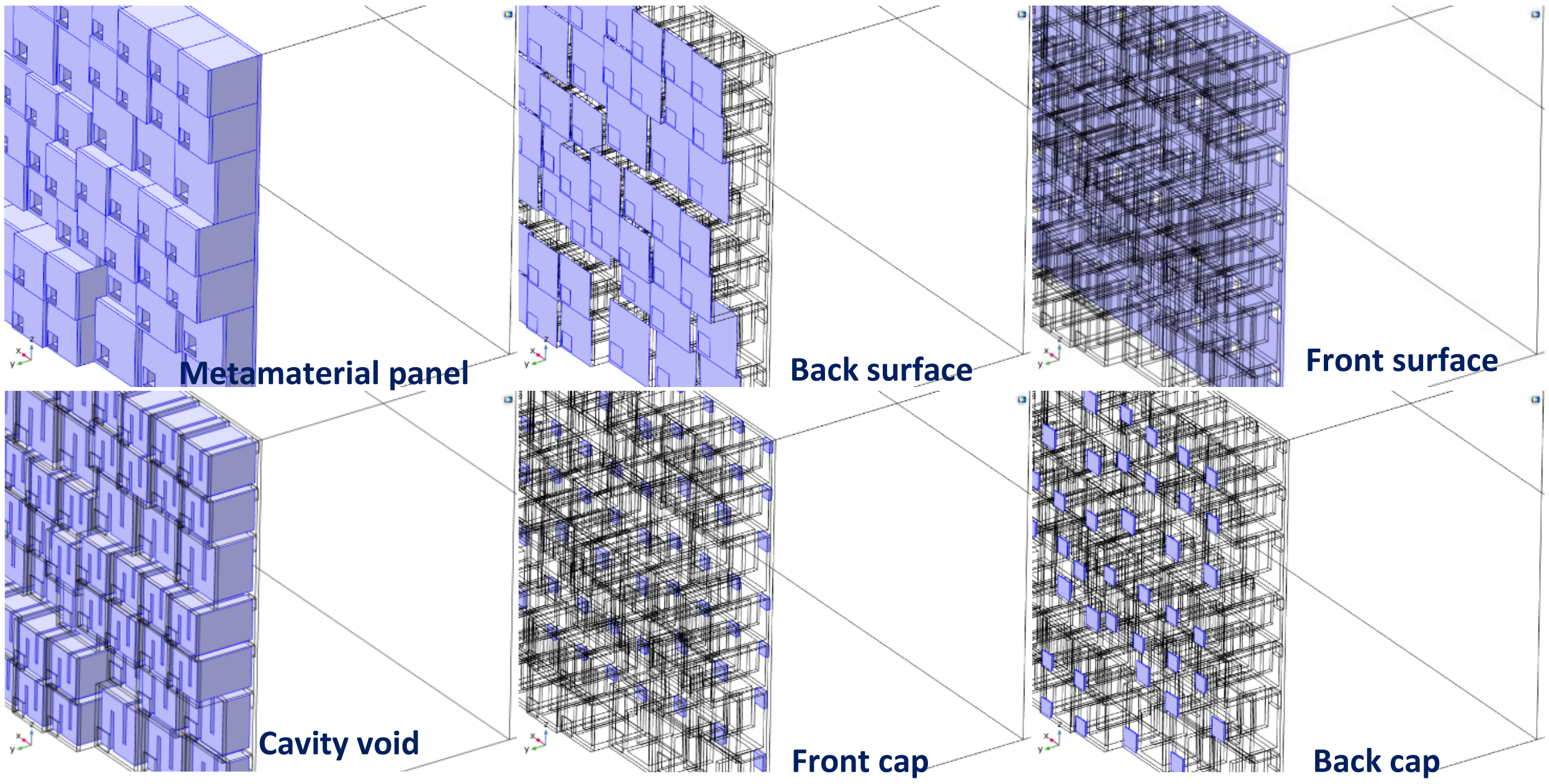

*Figure 8: Structural and acoustic domains in metamaterial panel macrocell simulations*

The full FE model is shown in Fig. 9. Periodic Bloch-Floquet boundary conditions are applied to the sides of the acoustic and structural domains. A source load of plane wave radiation is applied to the wave path with the same pressure field described in the previous paragraph. The structural domain is modelled using 278659 solid elements. The acoustic domains are modelled using 539926 elements. The maximum element size in the acoustic region and in the backing cavity region is $\Delta = \frac{\lambda_{min}}{10}$, where $\lambda_{min}$ is the smallest wavelength and decreases to $\Delta = \frac{\lambda_{min}}{30}$ in the AM region. Viscous-thermal losses are modelled using the "narrow region acoustics" model, which allows to account for the thermal and viscous losses that occur in narrow ducts where the cross-section is comparable in size to the thermal and viscous boundary layer thickness. This option allows to avoid solving a full thermos-viscous acoustic model in the whole domain. The narrow regions are defined in the front and back caps and in the UC cavities. The loss is calculated based on rectangular duct, taking the width (*w*) and the thickness (*t*) of the duct as parameters. In the configurations where the back caps are closed, no narrow regions are applied. Once again, structural domains are considered perfectly rigid, since results from fully vibro-acoustic and simple acoustic simulations bring about negligible differences.

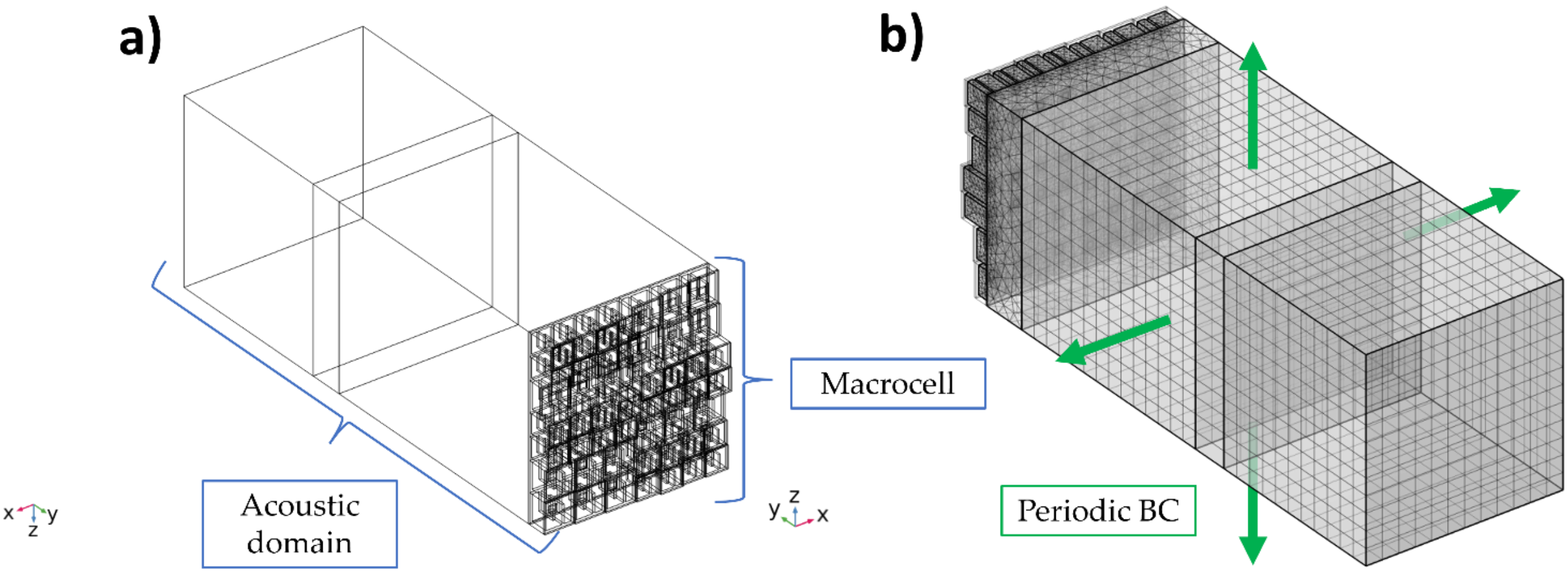

*Figure 9: FE model of the AM panel: a) Rectangular duct ("acoustic domain") in which acoustic wave propagation is simulated, with normal incidence on the labyrinthine AM panel and backing cavity; b) Corresponding FE mesh, with refinement in the AM region.*

### *4.3 AM panel characterization*

The measurement of the sound absorption of the complete panel is performed in a small-scale reverberation room at the Politecnico di Torino. A full validation of the performance of the room is presented in [21], highlighting the 400-5000 Hz general validity frequency range. The mean reverberation time of the empty room between 100 Hz and 5000 Hz is of 0.95 s with a Schroeder frequency $f_s$ of 1152 Hz [45]. In order to ensure a high diffusivity of the sound field, the room is equipped with 8 diffusers on the ceiling, covering 13.5% of the total room area. The panel characterization procedure consists in using the integrated impulse response method [41] for simultaneous measurements on six different microphone positions in two conditions, i.e. with and without the sample on the floor of the room. The measurement chain includes a set-up of six 1/4" BSWA Tech MPA451 microphones and ICP104 (BSWA Technology Co., Ltd., Beijing, China), two ITA High-Frequency Dodecahedron Loudspeakers with their specific ITA power amplifiers (ITA-RWTH, Aachen, Germany) and a Roland OctaCapture UA-1010 sound card (Roland Corporation, Japan) in order to perform 12 measurements (the minimum number required by ISO 354 [41]). We use a Matlab code combined with the functions of the ITA-Toolbox (an open-source toolbox from RWTH-Aachen, Germany) for sound generation, recording and signal processing. We perform a spatial averaging considering all the 12 sources and microphone combinations, checking the temperature (≥ 15 °C) and humidity (between 30-90 %) conditions. In accordance with ISO 354, before measurements the equivalent specimen absorption area is calculated (in square meters), using the formula

$$A_T = 55.3V\left(\frac{1}{c_2 t_2} - \frac{1}{c_1 t_1}\right) - 4V(m_2 - m_1) \quad (2)$$

where $t_1, t_2$ are the reverberation times (in s) of the chamber without and with the specimen, respectively; $V$ is the volume of the empty reverberation room in $m^3$; $c_1$ and $c_2$ are the sound propagation speeds in air in the reverberation room without and with the sample, respectively; $m_1, m_2$ are the power attenuation coefficients of the climatic conditions in the reverberation room without and with the sample. Finally, the random-incidence absorption coefficient of the tested specimen is computed as $\alpha = \frac{A_T}{S}$, where $S$ is the area covered by the test specimen, expressed in $m^2$.

### *4.4 AM panel sound absorption results*

As shown in Fig. 7, the panel has a flat upper surface and an uneven lower surface. The upper surface is equipped with apertures from which the sound waves access the UCs, while the lower surface is fully sealed. Fig. 10a displays the experimental test configuration of the panel: the flat surface is facing the sound source. This configuration corresponds to the configuration in which the panel has an attenuating effect on sound waves. In Fig. 10b, the comparison of analytically, numerically and experimentally predicted absorption spectra is shown. Frequencies of interest are reported as third-octave bands in the range of interest (250-5000 Hz). The orange line shows the effective sound absorption of the upper surface, which is close to the ideal value of 1 in the desired frequency range between 800 and 1300 Hz. Additionally, an analytical absorption curve (in blue) is obtained for a fine band spectrum computing the overall impedance of the panel considering the UCs working in parallel, as detailed in [37]. Although the pressure field of the analytical and numerical model impinges normally on the surface, while in experimental testing conditions of the reverberation room there is a diffused field, the two absorption spectra show good agreement, in the main absorption peak range, indicating the effectiveness of the panel also in non-normal conditions. Numerical results, calculated with the model described in the previous Section (again for normal incidence), are also included in the plot in Fig. 10b, showing excellent agreement with analytical and experimental results. Some discrepancy emerges at higher frequencies (above 3 kHz) between experimental and predicted results, probably due to the greater influence of a non-normal incidence angle.

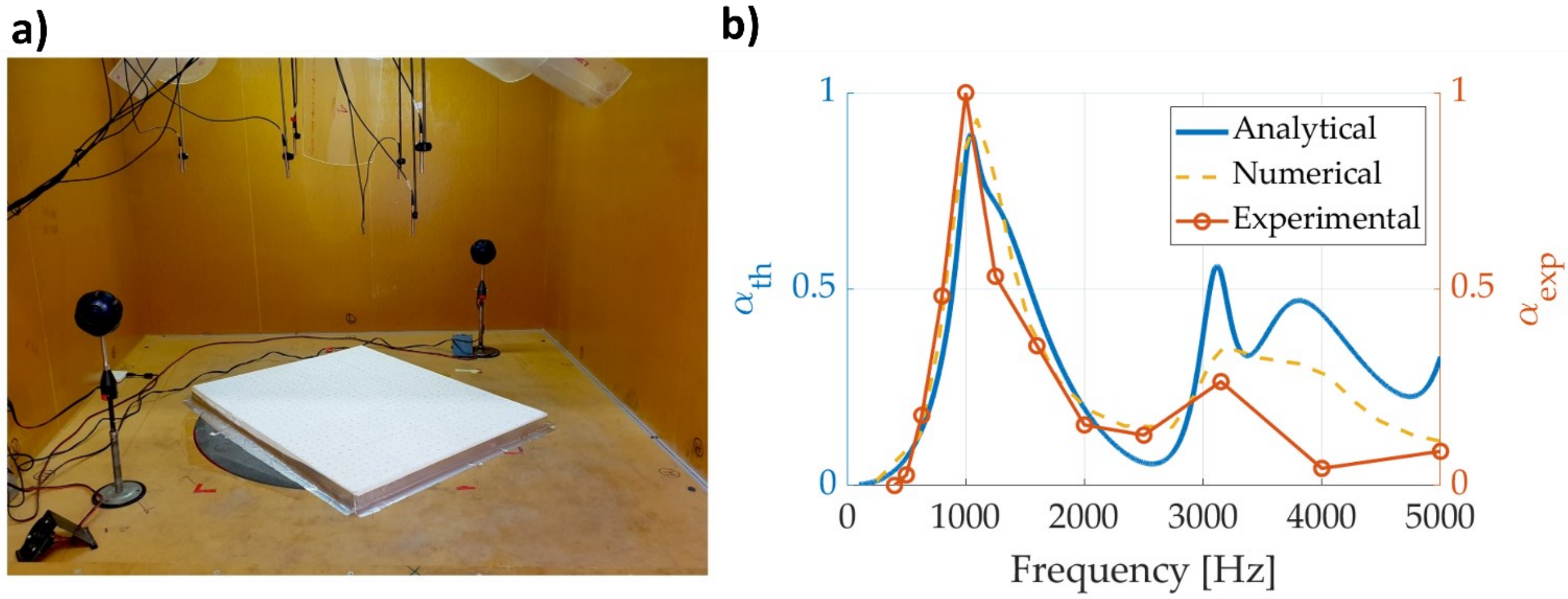

*Figure 10: a) AM panel absorption measurement setup in the small-scale reverberation room; b) Absorption spectrum of the labyrinthine AM panel in the reverberation room measurements. The experimental spectrum is compared to the analytically and numerically predicted ones.*

## 5 Numerical evaluation of different labyrinthine sound absorption panel solutions

### *5.1 Macrocell with backing cavity*

Given the promising results for the design procedure outlined in the previous Sections, and the good agreement between numerical and experimental results, a further numerical study is performed to evaluate possible developments of the AM-based sound absorption panel. The objective is to assess its acoustic performance under different design configurations. More specifically, three additional configurations are assessed: a) the addition of a melamine foam inside the UCs of the panel; b) the addition of a rigid rear backing cavity applied in correspondence with the uneven bottom surface of the panel (in this case with open apertures), with/without melamine foam filling; c) a combination of a) + b). The acoustic performance of each of these features can be studied and compared to the original configuration (taken as a baseline), again performing FE simulations with COMSOL Multiphysics, focusing on a single marcocell, taken as representative of the panel behaviour. Again, thermoviscous effects need to be accounted for.

In the second configuration, the AM panel is coupled to an additional acoustic domain of a depth equal to four times its side length (Fig.11a and b). Non-reflective Perfectly Matched Layer (PML) boundary conditions are applied to the back of the acoustic domain. Once again, structural domains are considered perfectly rigid, since results from fully vibro-acoustic and simple acoustic simulations give rise to negligible differences. The foam material is chosen as melanine, whose properties are summarized in Table 1.

*Table 1: Melamine foam material properties used in FE simulations.*

| Parameter | Value |
|---|---|
| Porosity | 0.995 |
| Flow resistivity | 10500 $Pa \cdot s/m^2$ |
| Thermal characteristic length | 470 $\mu$m |
| Viscous characteristic length | 240 $\mu$m |
| Tortuosity factor | 1.0059 |

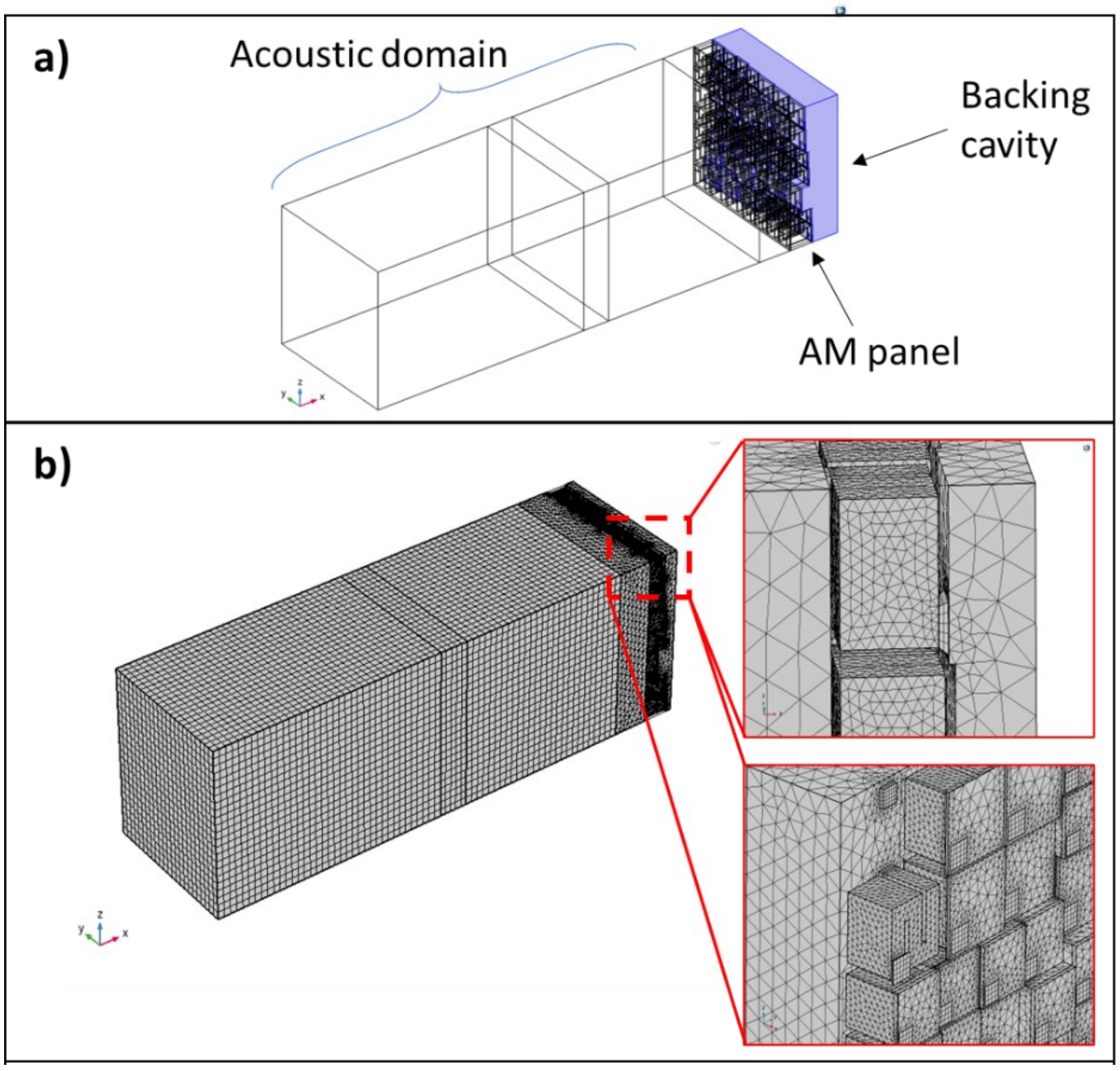

*Figure 11: FE model of the AM panel with a backing cavity: a) Rectangular duct ("acoustic domain") in which acoustic wave propagation is simulated, with normal incidence on the labyrinthine AM panel and backing cavity; b) Corresponding FE mesh, with refinement in the AM region, as shown in the side and rear view enlargements*

*5.2 Results*

Figure 12 shows the calculated absorption coefficient in the three considered cases. The baseline simulation, corresponding to the dashed curve in Fig. 10b, is shown in red. The effect of the presence of a backing cavity (5mm thick) filled with melamine foam (green dashedcurve in Fig.11) is to shift absorption peaks to lower frequencies (the lowest from 1 kHz to about 700 Hz), with a beneficial effect for the attenuation of low frequencies. The addition of a foam filling in the AM cavities, on the other hand, leads to a more uniform absorption coefficient over the whole frequency range, at the expense of a lower efficiency at the AM working frequencies. This solution can however be useful in the case where a more distributed attenuation effect is required over the whole frequency range, rather than at specific target frequencies. Other types of foams can also be considered to optimize the combined effect with the AM.

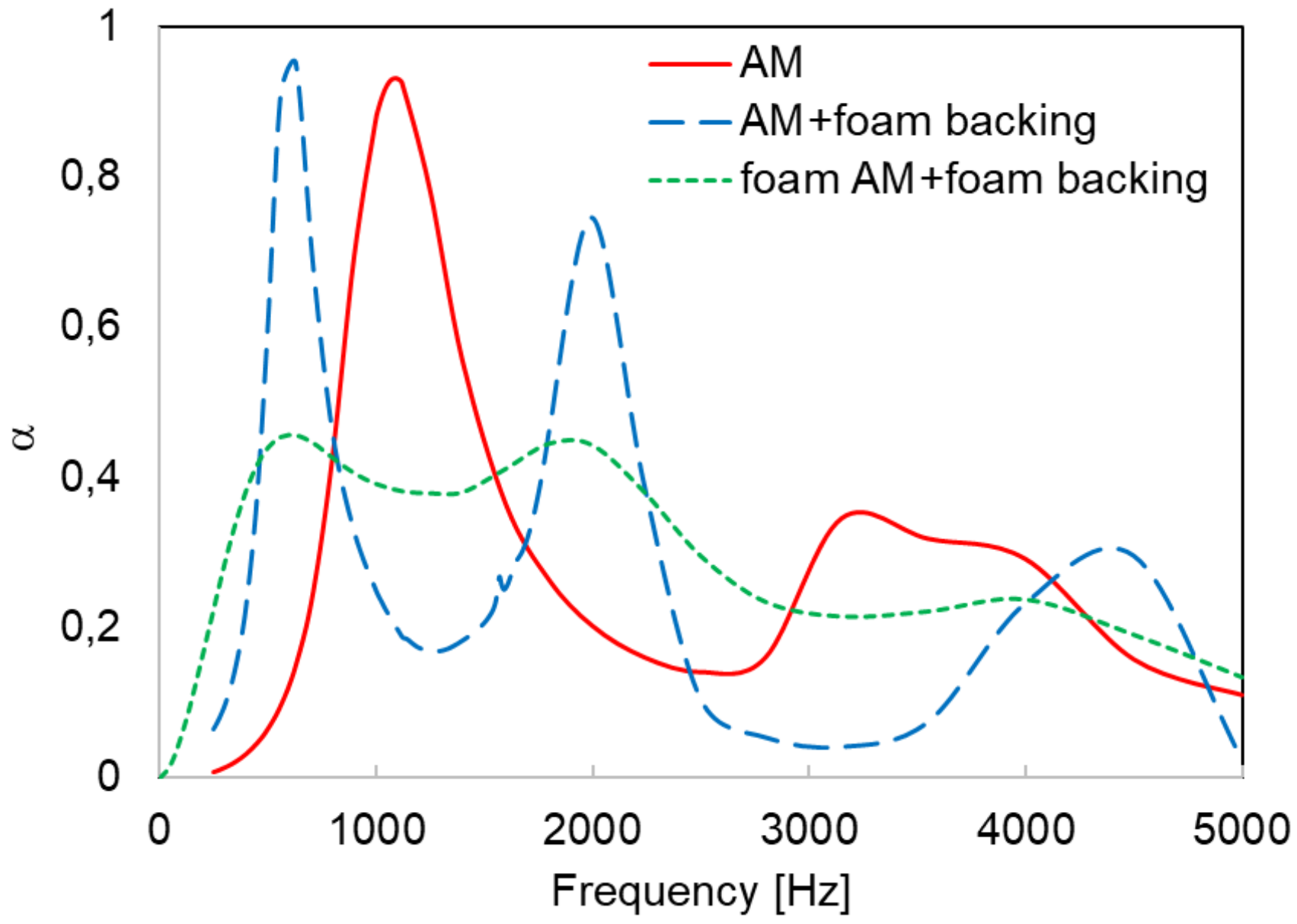

*Figure 12: FE simulation results for absorption coefficient vs. frequency for different sound attenuating panel solutions: the initial panel (AM), with the addition of a 5-mm foam-filled backing on the panel (AM+foam backing) and additionally filling the AM cavities with foam (foam AM+foam backing).*

## Conclusions

In this work, we have provided an experimental proof of concept for a novel approach to noise attenuation exploiting a rainbow-based design using labyrinthine metamaterials and combining UCs of varying thickness and lateral size in a quasi-periodic arrangement that ensures good homogeneity in the panel response. We have described the full design and validation procedure, from numerical design and modeling of the UC, to its characterization in an impedance tube, to the design of the macrocells composing the panel, and its realization using selective laser sintering. The final structure has then been characterized experimentally in a small-scale reverberation chamber, demonstrating a close to ideal absorption over the targeted low frequency range, centred at 1 kHz, thus validating the approach. Finally, detailed FE simulations have allowed to evaluate possible improvements/modifications to the panel by adding a foam filling and a foam backing cavity. The proposed prototype can be further developed, and thanks to its modular design, can be employed in diverse applications, e.g. in room acoustics, in automotive parts, or aeronautics in general. This approach, together with other metamaterial-based solutions proposed in the literature, contributes to an alternative (or complementary) route to the use of traditional sound absorbing materials in noise control. The proposed solution can be particularly attractive due to the reduced thickness of the panels and relatively small density of the required parts emerging from the subwavelength nature of the labyrinthine metamaterials used in the design.

## Acknowledgements

FN, VHK, LB, EM, DP, MZ, ASG, LS, FB thank the Alta Scuola Politecnica project "MetaMAPP". FN, ASG, NMP and FB are supported by the EU H2020 FET Open "Boheme" grant No. 863179.

## References

[1] G. Ma and P. Sheng, *Acoustic Metamaterials: From Local Resonances to Broad Horizons*, Science Advances 2, e1501595 (2016).

[2] B. Assouar, B. Liang, Y. Wu, Y. Li, J. C. Cheng, and Y. Jing, *Acoustic Metasurfaces*, Nature Reviews Materials 3, 460–472 (2018).

[3] H. Ge, M. Yang, C. Ma, M. H. Lu, Y. F. Chen, N. Fang, and P. Sheng, *Breaking the Barriers: Advances in Acoustic Functional Materials*, National Science Review 5, 159–182 (2018).

[4] X. Zhang, Z. Qu, and H. Wang, *Engineering Acoustic Metamaterials for Sound Absorption: From Uniform to Gradient Structures*, iScience 23, 101110 (2020).

[5] S. Kumar and H. Lee, *The Present and Future Role of Acoustic Metamaterials for Architectural and Urban Noise Mitigations*, Acoustics 1, 590-607 (2019).

[6] J. Li, X. Wen, and P. Sheng, *Acoustic Metamaterials*, J Appl Phys 129, 171103 (2021).

[7] F. Langfeldt and W. Gleine, *Improved Sound Transmission Loss of Glass Wool with Acoustic Metamaterials*, in *Proceedings of the 26th International Congress on Sound and Vibration, ICSV 2019* (2019).

[8] G. Palma, H. Mao, L. Burghignoli, P. Göransson, and U. Iemma, *Acoustic Metamaterials in Aeronautics*, Applied Sciences, 8(6), 971 (2018).

[9] M. Yang and P. Sheng, *Sound Absorption Structures: From Porous Media to Acoustic Metamaterials*, Annual Review Material Research 47, 83–114 (2017).

[10] L. Zhao and S. Zhou, *Compact Acoustic Rainbow Trapping in a Bioinspired Spiral Array of Graded Locally Resonant Metamaterials*, Sensors 19, 788 (2019).

[11] M. Yang, G. Ma, Z. Yang, and P. Sheng, *Coupled Membranes with Doubly Negative Mass Density and Bulk Modulus*, Physical Review Letters 110, 134301 (2013).

[12] G. Ma, M. Yang, S. Xiao, Z. Yang, and P. Sheng, *Acoustic Metasurface with Hybrid Resonances*, Nature Materials 13, 873–878 (2014).

[13] V. Romero-García, G. Theocharis, O. Richoux, and V. Pagneux, *Use of Complex Frequency Plane to Design Broadband and Sub-Wavelength Absorbers*, Journal of the Acoustical Society of America 139, 3395–3403 (2016).

[14] N. Jiménez, W. Huang, V. Romero-García, V. Pagneux, and J. P. Groby, *Ultra-Thin Metamaterial for Perfect and Quasi-Omnidirectional Sound Absorption*, Applied Physics Letters 109, 121902 (2016).

[15] V. Romero-Garciá, G. Theocharis, O. Richoux, A. Merkel, V. Tournat, and V. Pagneux, *Perfect and Broadband Acoustic Absorption by Critically Coupled Sub-Wavelength Resonators*, Scientific Reports 6, 19519 (2016)

[16] N. Jiménez, V. Romero-García, V. Pagneux, and J.-P. Groby, *Rainbow-Trapping Absorbers: Broadband, Perfect and Asymmetric Sound Absorption by Subwavelength Panels for Transmission Problems*, Scientific Reports 7, 13595 (2017).

[17] J. Boulvert, J. Costa-Baptista, T. Cavalieri, V. Romero-García, G. Gabard, E. R. Fotsing, A. Ross, M. Perna, J. Mardjono, and J.-P. Groby, *Folded Metaporous Material for Sub-Wavelength and Broadband Perfect Sound Absorption*, Applied Physics Letters 117, 251902 (2020).

[18] J. Boulvert, T. Humbert, V. Romero-García, G. Gabard, E. R. Fotsing, A. Ross, J. Mardjono, and J.-P. Groby, *Perfect, Broadband, and Sub-Wavelength Absorption with Asymmetric Absorbers: Realization for Duct Acoustics with 3D Printed Porous Resonators*, Journal of Sound and Vibration **523**, 116687 (2022).

[19] Z. Liang and J. Li, *Extreme Acoustic Metamaterial by Coiling up Space*, Physical Review Letters 108, 114301 (2012).

[20] Z. Liang, T. Feng, S. Lok, F. Liu, K. B. Ng, C. H. Chan, J. Wang, S. Han, S. Lee, and J. Li, *Space-Coiling Metamaterials with Double Negativity and Conical Dispersion*, Scientific Reports 3, 1614 (2013).

[21] Y. Xie, A. Konneker, B. I. Popa, and S. A. Cummer, *Tapered Labyrinthine Acoustic Metamaterials for Broadband Impedance Matching*, Applied Physics Letters 103, 201906 (2013)

[22] S. K. Maurya, A. Pandey, S. Shukla, and S. Saxena, *Double Negativity in 3D Space Coiling Metamaterials*, Scientific Reports 6, 33683 (2016)

[23] Y. Xie, B. I. Popa, L. Zigoneanu, and S. A. Cummer, *Measurement of a Broadband Negative Index with Space-Coiling Acoustic Metamaterials*, Physical Review Letters 110, 175501 (2013).

[24] Y. Liu, W. Xu, M. Chen, T. Yang, K. Wang, X. Huang, H. Jiang, and Y. Wang, *Three-Dimensional Fractal Structure with Double Negative and Density-near-Zero Properties on a Subwavelength Scale*, Materials & Design 188, 108470 (2020).

[25] G. Y. Song, Q. Cheng, B. Huang, H. Y. Dong, and T. J. Cui, *Broadband Fractal Acoustic Metamaterials for Low-Frequency Sound Attenuation*, Applied Physics Letters 109, 131901 (2016).

[26] C. Zhang and X. Hu, *Three-Dimensional Single-Port Labyrinthine Acoustic Metamaterial: Perfect Absorption with Large Bandwidth and Tunability*, Physical Review Applied **6**, (2016).

[27] A. O. Krushynska, F. Bosia, M. Miniaci, and N. M. Pugno, *Spider Web-Structured Labyrinthine Acoustic Metamaterials for Low-Frequency Sound Control*, New Journal of Physics (2017).

[28] A. O. Krushynska, F. Bosia, and N. M. Pugno, *Labyrinthine Acoustic Metamaterials with Space-Coiling Channels for Low-Frequency Sound Control*, Acta Acustica United with Acustica 104, 200-210 (2018).

[29] M. Molerón, M. Serra-Garcia, and C. Daraio, *Visco-Thermal Effects in Acoustic Metamaterials: From Total Transmission to Total Reflection and High Absorption*, New Journal of Physics **18**, (2016).

[30] *Foam Core Impresses in Aircraft Study*, Reinforced Plastics 58, 31-33 (2014).

[31] S. Ghinet, P. Bouche, T. Padois, L. Pires, O. Doutres, T. C. Kone, K. Triki, F. Abdelkader, R. Panneton, and N. Atalla, *Experimental Validation of Acoustic Metamaterials Noise Attenuation Performance for Aircraft Cabin Applications*, in *Proceedings of 2020 International Congress on Noise Control Engineering,* INTER-NOISE 2020 (2020).

[32] T. Cox and P. d'Antonio, *Acoustic Absorbers and Diffusers: Theory, Design and Application* (CRC press, 2016).

[33] N. Jiménez, O. Umnova, and J.-P. Groby, *Acoustic Waves in Periodic Structures, Metamaterials, and Porous Media*, Ch. "The Transfer Matrix Method in Acoustics", Springer International Publishing, (2021).

[34] Vincent Laude, *Phononic Crystals* (de Gruyter, 2020).

[35] V. Romero-García, G. Theocharis, O. Richoux, and V. Pagneux, *Use of Complex Frequency Plane to Design Broadband and Sub-Wavelength Absorbers*, Journal of the Acoustical Society of America 139, 3395 (2016).

[36] A. O. Krushynska, F. Bosia, and N. M. Pugno, *Labyrinthine Acoustic Metamaterials with Space-Coiling Channels for Low-Frequency Sound Control*, Acta Acustica United with Acustica 104, (2018).

[37] A. Magnani, C. Marescotti, and F. Pompoli, *Acoustic Absorption Modeling of Single and Multiple Coiled-up Resonators*, Applied Acoustics 186, 108504 (2022).

[38] Y. Wang, H. Zhao, H. Yang, J. Zhong, D. Zhao, Z. Lu, and J. Wen, *A Tunable Sound-Absorbing Metamaterial Based on Coiled-up Space*, Journal of Applied Physics 123, 185109 (2018).

[39] G. do N. Almeida, E. F. Vergara, L. R. Barbosa, and R. Brum, *Low-Frequency Sound Absorption of a Metamaterial with Symmetrical-Coiled-up Spaces*, Applied Acoustics 172, 107593 (2021).

[40] M. R. Stinson, *The Propagation of Plane Sound Waves in Narrow and Wide Circular Tubes, and Generalization to Uniform Tubes of Arbitrary Cross-sectional Shape*, Journal of the Acoustical Society of America 89, 550 (1991).

[41] ISO 354, *Acoustics — Measurement of Sound Absorption in a Reverberation Room*, International Standard (2003).

[42] ASTM, *Standard Test Method for Impedance and Absorption of Acoustical Materials Using a Tube, Two Microphones and a Digital Frequency Analysis System*, American Society for Testing of Materials (1990).

[43] ASTM E2611, *Standard Test Method for Normal Incidence Determination of Porous Material Acoustical Properties Based on the Transfer Matrix Method E2611*, American Society for Testing of Materials (2019).

[44] D. Pilon, R. Panneton, and F. Sgard, *Behavioral Criterion Quantifying the Effects of Circumferential Air Gaps on Porous Materials in the Standing Wave Tube*, Journal of the Acoustical Society of America 116, 344–356 (2004).

[45] M. R. Schroeder, *The '"Schroeder Frequency"' Revisited*, Journal of the Acoustical Society of America 99, 3240–3241 (1996).

[46] G. Kirchhoff, *Ueber Den Einfluss Der Wärmeleitung in Einem Gase Auf Die Schallbewegung*, Annalen der Physik 210, 177 (1868).

[47] C. Zwikker and C. W. Kosten, *Sound Absorbing Materials*, Elsevier (1949).

[48] N. Atalla and F. Sgard, *Modeling of Perforated Plates and Screens Using Rigid Frame Porous Models*, Journal of Sound and Vibration 303, 195 (2007).

**Appendix I**

The general theory of sound propagation in a cylindric tube in presence of viscous-thermal losses was initially developed by Kirchhoff [46] using the linearized Navier-Stokes equation. Given the complexity of the analytical solution, Zwikker and Kosten [47] developed approximate solutions for narrow and wide tube diameters. Subsequently, Stinson [40] derived an approximate solution for an infinitely long tube with arbitrary cross-sectional shape. This model was employed to calculate the variation of impedance along the coiled-up tube (Fig. 1b). The thin inlet was instead described using the Johnson-Champoux-Allard model (JCA), as pointed out in [48]. The model allows to include not only the losses taking place along the inlet, but also those around the aperture due to the sudden cross-section variation.

Considering the rigid backing as a wall of infinite impedance we can write the impedance beneath the aperture ($Z\,(x_1^-)$) as:

$$Z\,(x_1^-) = -iZ_{eff}^{c} \cot\,(k_{eff}^{c} x_1), \tag{3}$$

where $x_1 = 3l_c$, $l_c = \sqrt{a^2 + w^2} - \psi$, and $x_1^-$ indicates the $x_1$ value inside the cavity. $\psi$ is an empirical correction factor that models the effective propagation path due to the initial assumptions of an infinite straight tube of the Stinson's model [30]. The comparison with experimental results have shown that this factor is approximately 9% or less of the UC size $a$. $Z_{eff}^{c} = \sqrt{\rho_{eff}^{c} / C_{eff}^{c}}$, and $k_{eff}^{c} = \omega\sqrt{\rho_{eff}^{c} C_{eff}^{c}}$ are the effective impedance and propagation constants, where $\omega$ is the angular frequency, and $\rho_{eff}^{c}$ and $C_{eff}^{c}$ are the complex density and compressibility function averaged over the channel cross-section according to the Stinson's approximation:

$$\rho_{eff}^c = \frac{\rho_0 \nu w^2 t^2}{4 i\omega}\left[\sum_{k=1}^{\infty}\sum_{n=1}^{\infty}\left[\alpha_k^2\,\beta_n^2\left(\alpha_k^2+\beta_n^2+\frac{i\omega}{\nu}\right)\right]^{-1}\right]^{-1}, \tag{4}$$

$$C_{eff}^c = \frac{1}{P_0}\left[1-\frac{4i\omega(\gamma-1)}{\nu' w^2 t^2}\sum_{k=1}^{\infty}\sum_{n=1}^{\infty}\left[\alpha_k^2\,\beta_n^2\left(\alpha_k^2+\beta_n^2+\frac{i\omega\gamma}{\nu'}\right)\right]^{-1}\right], \tag{5}$$

where $\alpha_m = \frac{\left(m+\frac{1}{2}\right)\pi}{w/2}$, $\beta_m = \frac{\left(m+\frac{1}{2}\right)\pi}{t/2}$, $\nu = \eta/\rho_0$ , $\eta$ is the air dynamic viscosity, and $\nu' = \kappa/\rho_0 C_v$ with $\kappa$ the thermal conductivity of air, $C_v$ the specific heat at constant volume of air and $\gamma = C_p/C_v$ where $C_p$ is the specific heat at constant pressure of air. The impedance at the aperture's top entrance ($Z\,(x_2^-)$) is computed as:

$$Z\,(x_2^-) = Z_{eff}^{\,a}\,\frac{-iZ\,(x_1^+)\cot\left(k_{eff}^a x_2\right)+Z_{eff}^{\,a}}{Z\,(x_1^+)-iZ_{eff}^{\,a}\cot\left(k_{eff}^a x_2\right)}, \tag{6}$$

where and $x_1^+$ indicates the $x_1$ value inside the aperture, $x_2^-$ indicates the $x_2$ value inside the aperture, $Z\,(x_1^+) = Z\,(x_1^-)\frac{S_A}{S_C}$, with $S_A = w^2$ and $S_C = w\times t$, $Z_{eff}^{\,a} = \sqrt{\rho_{eff}^a/C_{eff}^a}$, $k_{eff}^a = \omega\sqrt{\rho_{eff}^a\,C_{eff}^a}$ are the effective impedance and propagation constant for the aperture and where $\rho_{eff}^a$ and $C_{eff}^a$ are the complex density and compressibility functions at the aperture according to the JCA model [37]:

$$\rho_{eff}^a = \frac{\alpha_\infty\rho_0}{\Phi}+\frac{\sigma}{i\omega}\sqrt{1+\frac{4i\alpha_\infty^2\eta\rho_0\omega}{\sigma^2\Lambda^2\Phi^2}}, \tag{7}$$

$$C_{eff}^a = \frac{\gamma-(\gamma-1)\left[1+\frac{8\eta}{i\rho_0\omega Pr\Lambda'^2}\sqrt{1+\frac{i\rho_0\omega Pr\Lambda'^2}{16\eta}}\right]^{-1}}{\gamma P_0/\Phi}, \tag{8}$$

where $\alpha_\infty = 1$ is the tortuosity of the aperture, $\Phi = 1$ is the porosity of the panel, $\Lambda$ and $\Lambda'$ are the viscous and thermal characteristic lengths, which in this case are both equal to the hydraulic radius of the aperture, here computed as $\mathrm{r_{hyd}} = w/4$. $\sigma = \left[\left(\frac{2h}{r_{hyd}}+1\right)\ (\alpha_s R_s/\Phi)\ (1/\mathrm{h})\right] S_A/S_T$, where $\alpha_s$ is an empirical fitting parameter accounting for the viscous effects occurring at sharp edges [37] and $R_s = \frac{1}{2}\sqrt{2\eta\omega\rho_0}$ is the specific airflow resistance.

Finally, the impedance at the surface is computed as:

$$Z_S = Z(x_2^+) = Z(x_2^-)\frac{S_T}{S_A}, \tag{9}$$

Where $x_2^+$ indicates the $x_2$ value inside the tube, and $S_T$ is the cross section area of the.

Kundt's Tube (diameter $\varphi_0 = 35$ mm) in which the relative sample is experimentally measured. The analytical absorption spectrum (Fig. 2a) obtained setting $\psi = 0$, and $\alpha_s$=1, shows a perfect absorption peak attained at $1370$ Hz, i.e., for a normal incident acoustic wave of wavelength $\lambda \sim 250$ mm, which is about $26$ times larger than the UC thickness.